\documentclass[twoside,twocolumn,9pt]{article}
\usepackage{extsizes}
\usepackage[super,sort&compress,comma]{natbib} 
\usepackage[version=3]{mhchem}
\usepackage[left=1.5cm, right=1.5cm, top=1.785cm, bottom=2.0cm]{geometry}
\usepackage{balance}
\usepackage{mathptmx}
\usepackage{sectsty}
\usepackage{graphicx} 
\usepackage{lastpage}
\usepackage[format=plain,justification=justified,singlelinecheck=false,font={stretch=1.125,small,sf},labelfont=bf,labelsep=space]{caption}
\usepackage{float}
\usepackage{fancyhdr}
\usepackage{fnpos}
\usepackage[english]{babel}
\addto{\captionsenglish}{%
  
}
\usepackage{array}
\usepackage{droidsans}
\usepackage{charter}
\usepackage[T1]{fontenc}
\usepackage[usenames,dvipsnames]{xcolor}
\usepackage{setspace}
\usepackage[compact]{titlesec}
\usepackage{hyperref}
\usepackage{subfigure}

\usepackage{algorithmicx}
\usepackage{amsbsy}
\usepackage{amssymb}
\usepackage{amsmath}
\usepackage{booktabs}
\usepackage{siunitx}
\usepackage{makecell}
\usepackage{multirow}
\usepackage[ruled,vlined,linesnumbered]{algorithm2e}
\usepackage{mathrsfs}
\usepackage{xfrac}
\usepackage{soul}
\usepackage{bm}
\usepackage{multirow}
\usepackage{physics}
\usepackage{soul}

\usepackage{epstopdf}

\newcommand{\bld}{\bm}
\newcommand{\HF}{\mathrm{HF}}
\newcommand{\CC}{\mathrm{CC}}

\newcommand{\hf}{\mathrm{HF}}
\newcommand{\da}{\mathbf{D}^A}
\newcommand{\db}{\mathbf{D}^B}
\newcommand{\dmo}{\mathbf{D}}
\newcommand{\hmo}{\mathbf{h}}
\newcommand{\gmo}{\mathbf{G}}

\newcommand{\pota}{\mathbf{V}^A}
\newcommand{\potb}{\mathbf{V}^B}
\newcommand{\ha}{\mathbf{h}^A}
\newcommand{\hb}{\mathbf{h}^B}

\definecolor{cream}{RGB}{222,217,201}

\begin{document}

\pagestyle{fancy}
\thispagestyle{plain}
\fancypagestyle{plain}{
\renewcommand{\headrulewidth}{0pt}
}

\makeFNbottom
\makeatletter
\renewcommand\LARGE{\@setfontsize\LARGE{15pt}{17}}
\renewcommand\Large{\@setfontsize\Large{12pt}{14}}
\renewcommand\large{\@setfontsize\large{10pt}{12}}
\renewcommand\footnotesize{\@setfontsize\footnotesize{7pt}{10}}
\makeatother

\renewcommand{\thefootnote}{\fnsymbol{footnote}}
\renewcommand\footnoterule{\vspace*{1pt}%
\color{cream}\hrule width 3.5in height 0.4pt \color{black}\vspace*{5pt}} 
\setcounter{secnumdepth}{5}

\makeatletter 
\renewcommand\@biblabel[1]{#1}            
\renewcommand\@makefntext[1]%
{\noindent\makebox[0pt][r]{\@thefnmark\,}#1}
\makeatother 
\renewcommand{\figurename}{\small{Fig.}~}
\sectionfont{\sffamily\Large}
\subsectionfont{\normalsize}
\subsubsectionfont{\bf}
\setstretch{1.125} 
\setlength{\skip\footins}{0.8cm}
\setlength{\footnotesep}{0.25cm}
\setlength{\jot}{10pt}
\titlespacing*{\section}{0pt}{4pt}{4pt}
\titlespacing*{\subsection}{0pt}{15pt}{1pt}

\fancyfoot{}
\fancyfoot[LO,RE]{\vspace{-7.1pt}\includegraphics[height=9pt]{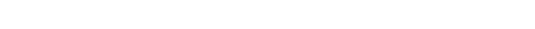}}
\fancyfoot[CO]{\vspace{-7.1pt}\hspace{11.9cm}\includegraphics{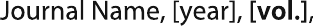}}
\fancyfoot[CE]{\vspace{-7.2pt}\hspace{-13.2cm}\includegraphics{RF}}
\fancyfoot[RO]{\footnotesize{\sffamily{1--\pageref{LastPage} ~\textbar  \hspace{2pt}\thepage}}}
\fancyfoot[LE]{\footnotesize{\sffamily{\thepage~\textbar\hspace{4.65cm} 1--\pageref{LastPage}}}}
\fancyhead{}
\renewcommand{\headrulewidth}{0pt} 
\renewcommand{\footrulewidth}{0pt}
\setlength{\arrayrulewidth}{1pt}
\setlength{\columnsep}{6.5mm}
\setlength\bibsep{1pt}

\makeatletter 
\newlength{\figrulesep} 
\setlength{\figrulesep}{0.5\textfloatsep} 

\newcommand{\topfigrule}{\vspace*{-1pt}%
\noindent{\color{cream}\rule[-\figrulesep]{\columnwidth}{1.5pt}} }

\newcommand{\botfigrule}{\vspace*{-2pt}%
\noindent{\color{cream}\rule[\figrulesep]{\columnwidth}{1.5pt}} }

\newcommand{\dblfigrule}{\vspace*{-1pt}%
\noindent{\color{cream}\rule[-\figrulesep]{\textwidth}{1.5pt}} }

\makeatother

\twocolumn[
  \begin{@twocolumnfalse}
{\includegraphics[height=30pt]{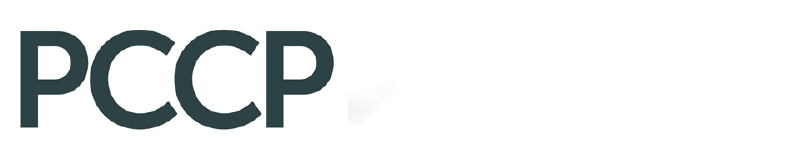}\hfill\raisebox{0pt}[0pt][0pt]{\includegraphics[height=55pt]{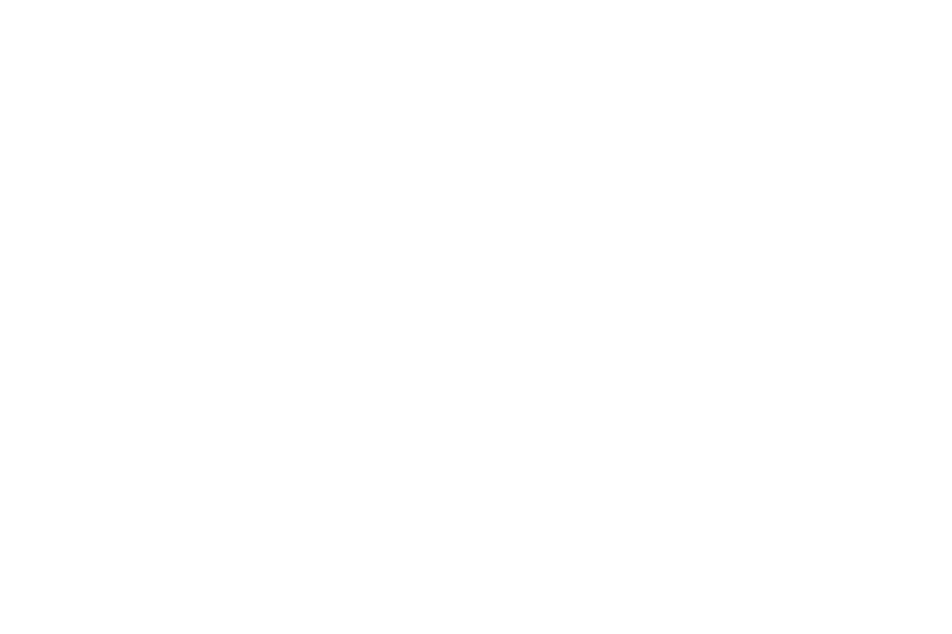}}\\[1ex]
\includegraphics[width=18.5cm]{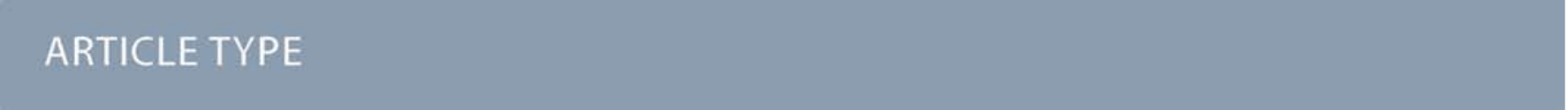}}\par
\vspace{1em}
\sffamily
\begin{tabular}{m{4.5cm} p{13.5cm} }

\includegraphics{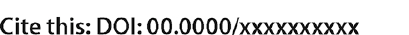} & \noindent\LARGE{\textbf{Linear Response Properties of Solvated Systems: A Computational Study$^\dag$}} \\
\vspace{0.3cm} & \vspace{0.3cm} \\

 & \noindent\large{Linda Goletto,\textit{$^{a\ddag}$} Sara Gómez,\textit{$^{b}$} Josefine H. Andersen,\textit{$^{c}$} 
 Henrik Koch,\textit{$^{a,b}$}$^{\ast}$ and Tommaso Giovannini\textit{$^{b}$}$^{\ast}$} \\

\includegraphics{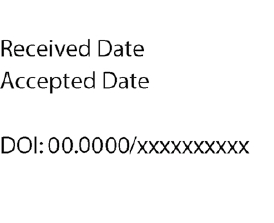} & \noindent\normalsize{We present a computational study of static and dynamic linear polarizabilities in solution. We use different theoretical approaches to describe solvent effects, ranging from quantum mechanics/molecular mechanics (QM/MM) to quantum embedding approaches. In particular, we consider non-polarizable and polarizable QM/MM methods, the latter based on the fluctuating charge (FQ) force field. In addition, we use a quantum embedding method defined in the context of multilevel Hartree-Fock (MLHF), where the system is divided into active and inactive regions, and combine it with a third layer described by means of the FQ model. The multiscale approaches are then used as reference wave functions for equation-of-motion coupled cluster (EOM-CC) response properties, allowing for the account of electron correlation. The developed models are applied to the calculation of linear response properties of two organic moieties---namely, para-nitroaniline and benzonitrile---in non-aqueous solvents---1,4-dioxane, acetonitrile, and tetrahydrofuran. The computed polarizabilities are then discussed in terms of the physico-chemical solute-solvent interactions described by each method (electrostatic, polarization and Pauli repulsion), and finally compared with the available experimental references.
} \\

\end{tabular}

\end{@twocolumnfalse} \vspace{0.6cm}

]


\renewcommand*\rmdefault{bch}\normalfont\upshape
\rmfamily
\section*{}
\vspace{-1cm}


\footnotetext{\textit{$^{a}$~Department of Chemistry, Norwegian University of Science and Technology (NTNU), NO-7491 Trondheim, Norway}}
\footnotetext{\textit{$^{b}$~Scuola Normale Superiore, Piazza dei Cavalieri 7, IT-56126 Pisa, PI Italy; E-mail: tommaso.giovannini@sns.it; henrik.koch@sns.it}}
\footnotetext{\textit{$^{c}$~DTU Chemistry, Technical University of Denmark, DK-2800 Kongens Lyngby, Denmark}}
\footnotetext{\textit{$^{\ddag}$}~Present address: \textit{Scuola Normale Superiore, Piazza dei Cavalieri 7, I-56126 Pisa, Italy}}

\footnotetext{\dag~Electronic Supplementary Information (ESI) available: data related to Fig.~\ref{fig:pna_dip}-\ref{fig:bnz-in-thf_plots}. See DOI: 10.1039/cXCP00000x/}



\section{Introduction}
The response of a molecular system to an external electric field plays a fundamental role in a plethora of technological applications.\cite{helgaker2012recent} In this context, theoretical chemistry can help understand the underlying physics of the different phenomena. Among them, linear response properties are the most basic quantities to be investigated, being the physico-chemical foundation of many different spectroscopic signals.\cite{norman2018principles,amos.1986,jensen2009atomistic,norman1997ab,norman2003polarization} Therefore, the theoretical modeling can have a pivotal role in gaining insight into how molecules and complex systems behave in the presence of electromagnetic radiation.\cite{norman2018principles} Of particular interest are molecular systems embedded in an external environment, being it a solvent or a biological matrix.\cite{cammi2000attempt,pedersen2014damped} In fact, in such cases, the molecular properties of the chromophore, which is usually the target of the study, can be drastically perturbed by the presence of the environment.\cite{egidi2014stereoelectronic} 

To face this kind of problems, the most widespread approach is to resort to the so-called focused models.\cite{warshel1976,tomasi2005} In the specific case of solutions, a high level of theory is used for the region of interest---the solute---while the solvent is treated at a less sophisticated level.\cite{mennucci2019multiscale} Thus, within the focused model formalism, the system is usually partitioned into two layers. In particular, the solute is generally described at the quantum mechanical (QM) level, while the treatment of the solvent can range from a continuum,\cite{tomasi2005} to molecular mechanics (MM) approaches,\cite{warshel1976, lin2007qm, senn2009} to a lower-level QM model.\cite{marrazzini2021multilevel,saether2017,svensson1996oniom,govind1999electronic,wesolowski2015,sun2016quantum,olsen2015polarizable} Each of these subcategories encompasses many different methods, where different solute-solvent interactions---e.g. electrostatic effects, polarization, Pauli repulsion, dispersion---are taken into account, and different computational efforts are required. In order to correctly describe strong and specific solute-solvent interactions, the atomistic nature of the environment usually needs to be retained in the modeling.\cite{giovannini2020csr} Among the atomistic models, the most commonly used belong to the family of QM/MM approaches.\cite{senn2009qm} In their basic formulation---the so-called electrostatic embedding---the environment electrostatically perturbs the QM density, but not vice versa.\cite{senn2009qm} To include mutual solute-solvent polarization effects, which might have a huge influence on the QM properties and spectra, polarizable QM/MM approaches are exploited.\cite{curutchet2009electronic,bondanza2020polarizable,olsen2011molecular,olsen2010excited} In this way, both electrostatic and polarization contributions are taken into account. However, non-electrostatic interactions such as Pauli repulsion and dispersion are usually neglected, although they play a crucial role in many complex systems.\cite{giovannini2017disrep,slipchenko2016effective,giovannini2019quantum} To recover a theoretically consistent picture of such interactions, which are intrinsically of quantum nature, quantum embedding approaches can be used.\cite{marrazzini2021multilevel,saether2017,svensson1996oniom,govind1999electronic,wesolowski2015,sun2016quantum,olsen2015polarizable,reinholdt2017polarizable} As mentioned above, these models are based on the description at the QM level---although less sophisticated than the one used for the solute---of at least a part of the environment. This allows for the treatment of solute-solvent Pauli repulsion, and in some cases of dispersion interactions too.\cite{folkestad2019multilevel,myhre2014multi,myhre2016multilevel} Due to the quantum description of a larger part of the system, quantum embedding approaches are generally more computationally demanding than QM/MM methods. This problem can be solved by three-layer approaches, where the largest part of the environment, usually the farthest from the solute, is described by means of classical force fields.\cite{wanko2008effect,olsen2015polarizable, bennie2016projector, nogueira2018effect, macetti2021three,goletto2021combining} In this way, within a small portion of the system most interactions are treated at the QM level, whereas long-range contributions are retained at the classical level only, providing a physically consistent picture. 
In this work, we present a computational investigation of linear response properties of two organic systems, namely para-nitroaniline (PNA) and benzonitrile (PhCN), dissolved in dioxane (DIO), acetonitrile (ACN), and tetrahydrofuran (THF). To quantify the solvent effects on such properties, we present a hierarchy of solvation approaches, ranging from common QM/MM methods to three-layer quantum embedding models.
As for QM/MM approaches, we consider both non-polarizable and polarizable frameworks.
The latter is based on the fluctuating charge (FQ) force field,\cite{cappelli2016integrated,giovannini2020csr,giovannini2020pccp} which has been recently parametrized for the selected solvents.\cite{ambrosetti2021quantum}
The three-layer quantum embedding, on the other hand, is based on the multilevel Hartree-Fock (MLHF) method.\cite{saether2017}
Within MLHF, the molecular orbitals (MOs) are partitioned into \textit{active} and \textit{inactive} by means of Cholesky decomposition\cite{beebe1977simplifications, sanchez2010,aquilante2006fast,folkestad2019efficient} coupled with projected atomic orbitals\cite{pulay1983, saebo1993} (PAOs) for the virtual space. The computational advantage of such a method lies in the fact that the active MOs are optimized in the field of the inactive ones, which are kept frozen, but orthogonal to the active space. Therefore, electrostatic and Pauli repulsion (and part of the polarization) active-inactive interactions are automatically taken into account at the HF level. To refine the picture provided by the basic formulation of MLHF, the active and inactive orbitals can be localized in their pre-defined spatial regions by means of an energy-based procedure that we have recently presented.\cite{giovannini2021energy} The model obtained is called MLHF-AB. If such a procedure is applied to an HF optimized wave function, fully accounting for solute-solvent interactions, the resulting MOs are denoted as fragment localized MOs (FLMOs).\cite{giovannini2022fragment} To minimize its computational cost, MLHF(-AB) can be coupled to an external MM layer (MLHF(-AB)/MM).\cite{goletto2021combining} 

To calculate the linear response properties, we use the aforementioned two- and three-layer wave functions as the reference for a post-HF description of the solute. In fact, electron correlation has been proven particularly significant for the accurate modeling of both static and dynamic (hyper)polarizabilities.\cite{champagne2005basis, pecul2005density,christiansen1999coupled, kongsted2002qm, kongsted2003linear, hrsak2018polarizable} If the ground state is dominated by a single-determinant wave function, the coupled cluster (CC) hierarchy of methods arguably provides one of the most sophisticated descriptions of electron correlation.\cite{pinkbook} For this reason, coupled cluster is often considered the theoretical golden standard for the prediction of linear response properties, although many other \emph{ab-initio} methods, ranging from density functional theory (DFT) to Møller-Plesset (MP) perturbation theory, have been routinely used for this purpose.\cite{limacher2009accurate, baranowska2013performance, wormer1986analysis, olsen1985linear} Note that, when dealing with excited states and molecular properties, coupled cluster methods typically follow one of two routes: response theory\cite{monkhorst1977calculation, koch1990coupled, pedersen1997coupled} (CCRT) or the equation-of-motion\cite{stanton1993equation, kobayashi1994calculation} (EOM-CC) formalism, which is exploited here. The two frameworks result in identical excitation energies, but differ in the molecular properties, although with a generally small discrepancy.\cite{helgaker2012recent}

The manuscript is organized as follows. In the next section, we detail the theoretical approaches with a focus on the solvation modeling and the calculation of linear response properties at the EOM-CC level of theory. Then, the computational protocol followed in the numerical analysis is presented and applied to the calculation of static and dynamic polarizabilities of PNA and PhCN dissolved in DIO, ACN, and THF. A summary and future perspectives end the manuscript.


\section{Theory}

This section outlines the theoretical basis of the solvation methods employed to compute the polarizabilities at the EOM-CC2 and EOM-CCSD levels, which are also briefly described. In particular, we briefly recall the theory of non-polarizable QM/MM and polarizable QM/FQ, together with that of the three-layer MLHF-AB/MM method.

\subsection{Non-polarizable QM/MM and polarizable QM/FQ}

As stated above, QM/MM methods rely on the partitioning of the total energy of the system into a QM ($E_{\text{QM}}$) and an MM ($E_{\text{MM}}$) contribution\cite{senn2009qm}
\begin{equation}
    E = E_{\text{QM}} + E_{\text{MM}} + E^{\text{int}}_{\text{QM/MM}}\ ,
\end{equation}
where $E^{\text{int}}_{\text{QM/MM}}$ is the QM/MM interaction energy. In electrostatic embedding, $E^{\text{int}}_{\text{QM/MM}}$ is limited to the purely electrostatic interaction, whereas in polarizable embedding, a polarization contribution is also included. In the former, each MM atom is endowed with a fixed charge. In the latter, if the FQ force field is exploited, the charge assigned to each atom can vary as a response to the QM potential. Since both the non-polarizable QM/MM and polarizable QM/FQ depend only on charges, the QM/MM interaction energy reads\cite{giovannini2020csr}
\begin{equation}
\label{eq:qmtip3p}
E^{\text{int}}_{\text{QM/MM}} = \sum_i q_i V_i(\mathbf{D})\ ,
\end{equation} 
where $V_i(\mathbf{D})$ is the QM potential due to the QM part acting on the i$th$ MM charge ($q_i$). While in electrostatic embedding such charges are fixed, in QM/FQ their values are obtained by solving the following set of linear equations\cite{giovannini2020csr,giovannini2020pccp} 
\begin{equation}
\label{eq:linearqmfq}
\mathbf{M}\mathbf{q}_{\lambda} = -\mathbf{C}_Q - \mathbf{V}(\mathbf{D}),
\end{equation}
where $\mathbf{q}_\lambda$ collects the FQ charges $q$ and suitable Lagrangian multipliers that ensure charge conservation and $\mathbf{M}$ is a matrix containing charge-charge interactions and Lagrangian blocks. The right hand side consists instead of $\mathbf{C}_Q$, which takes into account atomic electronegativities and the total charge constraints, and $\mathbf{V}(\mathbf{D})$---the QM potential. %
In both QM/MM approaches, the QM Fock matrix $F_{\mu\nu}$ (in the atomic orbital basis $\{\chi_\mu\}$) is modified by the inclusion of the QM/MM interaction\cite{giovannini2020csr}
\begin{equation}
    F_{\mu\nu} = h_{\mu\nu} + G_{\mu\nu}(\mathbf{D}) + \sum_i q_i V_{i,\mu\nu},
    \label{eq:fock-qmmm}
\end{equation}
where $h_{\mu\nu}$ and $G_{\mu\nu}$ are the one- and two-electron matrix elements. This additional term is fixed in non-polarizable QM/MM, whereas it varies at each self consistent field (SCF) step in QM/FQ, because the charges $q$ depend on the QM density.

\subsection{Multilevel Hartree-Fock}

In the MLHF model,\cite{saether2017} the total density matrix ($\dmo$) of the system is decomposed into an active ($\da$) and an inactive ($\db$) component. Under this assumption, the total energy of a system described at the HF level can be written as
\begin{equation}
\begin{aligned}
    E^{\mathrm{TOT}} = &\tr\hmo\da + \frac{1}{2}\tr\da\gmo(\da) + \tr\da\gmo(\db) \\ 
    + &\tr\hmo\db + \frac{1}{2}\tr \db\gmo(\db) + h_{\text{nuc}},
\label{eq:mlhf_en}
\end{aligned}
\end{equation}
where $\hmo$ and $\gmo$ are the one- and two-electron matrices, and $h_{\text{nuc}}$ is the nuclear repulsion energy. In MLHF, only $\da$ is iteratively optimized, whereas $\db$ is kept fixed during the SCF optimization. Thus, the last three terms in Eq. \ref{eq:mlhf_en}, i.e. the inactive energy and the nuclear repulsion, are constant throughout the procedure. Additionally, the minimization is performed in the MO space of the active part only, thus intrinsically reducing the computational cost of a full HF description. As a consequence, the MLHF Fock matrix elements take the following form
\begin{equation}
     F_{\mu\nu} = h_{\mu\nu} + G_{\mu\nu}(\mathbf{D}^A) + G_{\mu\nu}(\mathbf{D}^B),
    \label{eq:mlhf_f}
\end{equation}
where $G_{\mu\nu}(\mathbf{D}^B)$ describes the interaction between the active and inactive parts, and is indeed a one-electron term in the Fock matrix, because $\mathbf{D}^B$ is fixed. Within the MLHF framework the electrostatic, Pauli repulsion, and part of the polarization contributions between the active and inactive parts are described at the HF level. 

To further reduce the computational cost associated with an MLHF description, MLHF can be coupled with an additional MM layer, yielding the MLHF/MM method introduced in Ref. \citenum{goletto2021combining}. Between the MLHF and MM parts, the interaction is described at the purely electrostatic level, as in Eq. \ref{eq:qmtip3p}. In this case, $\mathbf{D}$ is obtained as the sum of the active and inactive density matrices. The MM layer can be equivalently treated at the non-polarizable or polarizable FQ level. In the latter case, in Eq. \ref{eq:linearqmfq} $\mathbf{D}$ refers to the total density matrix ($\da + \db$). Finally, the MLHF Fock matrix is modified by the coupling with the external MM layer as
\begin{equation}
     F_{\mu\nu} = h_{\mu\nu} + G_{\mu\nu}(\mathbf{D}^A) + G_{\mu\nu}(\mathbf{D}^B) + \sum_i q_i V_{i,\mu\nu}.
    \label{eq:mlhf_mm_f}
\end{equation}

After converging the MLHF(/MM) wave function, the active and inactive MOs can be localized in their specific spatial regions by using an energy-based localization of the MOs. In the resulting MLHF-AB approach,\cite{giovannini2021energy} the $\hmo$ contribution is separated in terms of the kinetic operator and the $A$, $B$, and interaction ($AB$) electron-nuclei potentials. Hence, the total energy can be rewritten as
\begin{equation}
\begin{aligned}
    E^{\mathrm{TOT}} & = \underbrace{\tr \ha\da + \frac{1}{2}\tr \da\gmo(\da) + h^{A}_{nuc}}_{E_A} \\
    & +\underbrace{\tr\hb\db + \frac{1}{2}\tr \db\gmo(\db) + h^{B}_{nuc}}_{E_B} \\
    & + \underbrace{\tr \potb\da + \tr\pota\db + \tr\da \gmo(\db) + h^{AB}_{nuc}}_{E_{AB}}.
\end{aligned}
\label{eq:mlhf-ab_en}
\end{equation}
The MOs of the A and B fragments are localized by means of a minimization of the $E_A + E_B$ energy in the space spanned by the occupied MOs of the two fragments. This means that the total density is not changed and that such a minimization is equivalent to a maximization of the repulsion energy $E_{AB}$. In this way, the occupied MOs of both fragments are localized in their specific spatial regions.
If a full HF optimization is performed before the localization procedure, FLMOs are obtained,\cite{giovannini2022fragment} yielding the HF\textsubscript{FLMOs} approach. Note that, if an additional MM layer is included in the modeling, it does not affect the minimization procedure: the total density matrix remains fixed, and so does the MLHF-MM interaction.
The localization procedure outlined here makes MLHF-AB a promising tool for the calculation of local properties, such as dipole moments\cite{giovannini2021energy} and polarizabilities. 

\subsection{EOM-CC2 and -CCSD linear polarizabilities}

After the SCF convergence is reached with any of the aforementioned approaches (QM/MM, MLHF-AB/MM, and HF$_{\text{FLMOs}}$/MM), the polarizabilities are computed at the EOM-CC2 and -CCSD level of theory by restraining the coupled cluster treatment to the QM (in QM/MM) or to the active part (in MLHF-AB/MM and HF$_{\text{FLMOs}}$/MM) only.

The coupled cluster wave function is expressed as the exponential parametrization\cite{pinkbook}
\begin{equation}
    \ket{\CC} = e^{T}\ket{\HF},
\end{equation}
where $\ket{\HF}$ is the reference Hartree-Fock wave function and $T$ is the cluster operator---i.e., the sum of all the excitation operators $\tau_{\nu}$ weighted by their amplitudes $t_{\nu}$
\begin{equation}
\begin{aligned}
    T =& T_1 + T_2 + ...\\
    =& \sum_{\nu_1} t_{\nu_1} \tau_{\nu_1} + \sum_{\nu_2} t_{\nu_2} \tau_{\nu_2} + ...
    \label{eq:t}
\end{aligned}
\end{equation}
Here, $\nu_n$ refers to the $n$-th electronic excitation. In the CC2\cite{christiansen1995second} and CCSD\cite{purvis1982full} models, the cluster operator $T$ is truncated after double excitations. The difference between the two models lies in the fact that in CC2 the double excitation component, $T_2$, is treated perturbatively. In CCSD, the doubles amplitudes are considered to infinite order, while the CC2 doubles are considered through first order only. The doubles amplitude equations of the two models (for a $T_1$-transformed Hamiltonian, $\bar{H}=e^{-T_1}He^{T_1}$) take the form\cite{christiansen1995second}
\begin{align}
    &\mathrm{CCSD:} \quad &\bra{\nu_2}\bar{H}+[\bar{H},T_2]+\frac{1}{2}[[\bar{H},T_2],T_2]\ket{\hf} = 0\\
    &\mathrm{CC2:}  \quad &\bra{\nu_2}\bar{H}+[F,T_2]\ket{\hf} = 0
\end{align}
while the singles equations remain the same for CC2 as in CCSD:
\begin{equation}
    \bra{\nu_1}\bar{H}+[\bar{H},T_2]\ket{\hf} = 0
\end{equation}
where
\begin{equation}
    \ket{\nu} = {\tau}_{\nu}\ket{\hf}, \hspace{1em}
    \bra{\nu} = \bra{\hf}{\tau}^{\dagger}_{\nu}.
\end{equation}
As a result, CC2 scales as $N^5$ compared to CCSD which is a $N^6$ model.

The CC2 model has a structure that is compatible with the exact linear response functions, and the CC2 response properties of a molecule can thus be computed within the same framework and formalism as CCSD.\cite{christiansen1995second}
Molecular response properties arise from the perturbation of an unperturbed system by an external time-periodic field, and can be expressed in terms of response functions. The second-order (linear) response to an external electric field  gives rise to the frequency-dependent electronic polarizability.

In the EOM-CC formalism,\cite{stanton1993coupled, stanton1993equation} the ground and excited states are explicitly parametrized as
\begin{equation}
    \ket{k} = e^{{T}} \sum_{\nu\ge0} R^k_{\nu} \ket{\nu}, \hspace{1em}
    \bra{\bar{k}} = \sum_{\nu\ge0}\bra{\nu}L^k_{\nu} e^{-{T}}.
    \label{eq:eom-es}
\end{equation}
Inserting the EOM states of eq.\ref{eq:eom-es} in the exact-state linear response function,\cite{linderberg2004propagators}
and considering the response of the components of the dipole moment operator $\mu$, the EOM-CC linear electronic polarizability, $\alpha$, reads\cite{kobayashi1994calculation, pedersen1997coupled}
\begin{equation}
    \alpha_{XY}(\omega) = \frac{1}{2}P^{XY}\left(\bld{\eta}^{X}\bld{t}^{Y}(\omega) + \bld{\eta}^{X}\bld{t}^{Y}(-\omega)\right),
\end{equation}
where the permutation operator $P^{XY}$ performs an interchange of the operators $X$ and $Y$, $\omega$ is the frequency of the external field, $X$ and $Y$ are cartesian components of $\mu$ and, for a generic operator $\cal{O}$,\cite{pawlowski2015response,faber2019rixseom}
\begin{align}
    \eta^{\cal{O}}_\nu &= 
    \left(\bra{\hf}+\bra{\bar{t}}\right) \bar{\cal{O}}\tau_\nu\ket{\hf} - \bar{t}_\nu\mel{\hf}{{\bar{\cal{O}}}}{\hf}\\
    \label{eq:eom-etax}
    &=
    \left(\bra{\hf}+\bra{\bar{t}}\right) [\bar{\cal{O}},\tau_\nu]\ket{\hf}
    + \sum_{\mu>\nu}\bar{t}_\mu\mel{\mu}{\tau_\nu\bar{\cal{O}}}{\hf} - \bar{t}_\nu(\bar{t}\,\xi^{\cal{O}}),
\end{align}
where we have introduced the similarity transformed operator $\bar{\cal{O}} = e^{-T}{\cal{O}} e^{T}$, and $\bra{\bar{t}} = \sum_\mu\bar{t}\bra{\mu}$ with $\bld{\bar{t}}$ being the ground state Lagrangian multipliers. 
The amplitude response vectors $\bld{t}^{\cal{O}}(\omega)$ are obtained by solving the linear equations
\begin{equation}
    (\mathbf{A} - \omega\mathbf{I})\bld{t}^{\cal{O}}(\omega) = - \bld{\xi}^{\cal{O}},
\end{equation}
where 
\begin{equation}
    \xi^{\cal{O}}_\nu = \bra{\nu} \bar{\cal{O}} \ket{\hf},
\end{equation}
and $\mathbf{A}$ is the \textit{coupled cluster Jacobian matrix} with elements
\begin{equation}
    A_{\nu\mu} = \bra{\nu} [\bar{H}, \hat{\tau}_{\mu}] \ket{\hf}.
\end{equation}


\section{Computational details}

\begin{figure}
    \centering
    \includegraphics[width=0.7\linewidth]{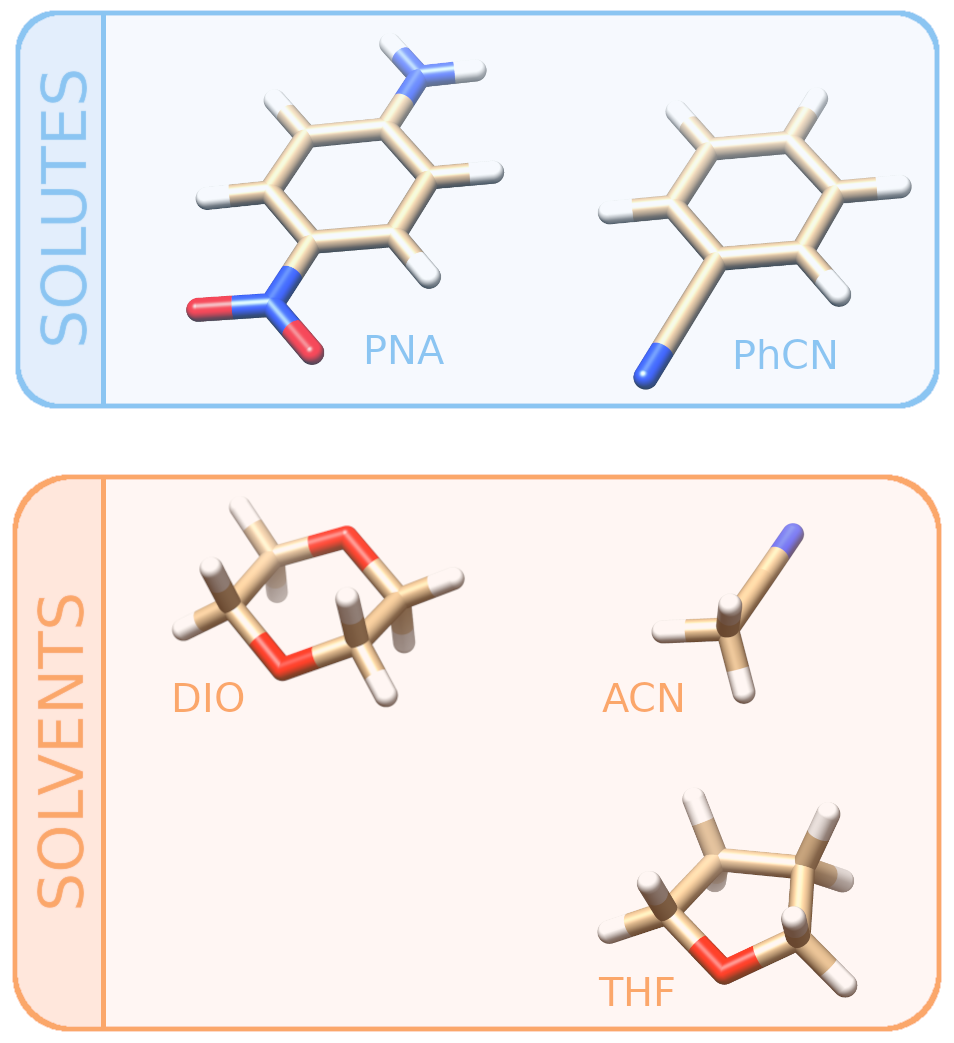}
    \caption{Molecular structures of the solutes (para-nitroaniline, PNA, and benzonitrile, PhCN) and solvents (1,4-dioxane, DIO, acetonitrile, ACN, and tetrahydrofuran, THF).}
    \label{fig:solutes-solvents}
\end{figure}

In this work, we select different organic molecules dissolved in non-aqueous environments, namely para-nitroaniline dissolved in 1,4-dioxane (PNA-in-DIO) and benzonitrile dissolved in both acetonitrile (PhCN-in-ACN) and tetrahydrofuran (PhCN-in-THF)---see Fig. \ref{fig:solutes-solvents}. Such systems have been selected because their measured linear polarizabilities have been previously reported in the literature.\cite{wortmann1993deviations, alvarado2003solvent}

In order to correctly sample the solute-solvent phase-space, classical molecular dynamics (MD) simulations are performed for both PNA and PhCN dissolved in the different environments. In the case of PNA-in-DIO, the MD simulation has been performed by following the procedure recently proposed in Ref. \citenum{ambrosetti2021quantum}. Similarly, for both PhCN-in-ACN and PhCN-in-THF, the General Amber Force Field (GAFF)\cite{wang2004development} is used to describe the solute and solvents, for which charges and parameters are obtained by using the RESP charge-fitting method\cite{bayly1993well} and the Antechamber package,\cite{wang2006automatic} respectively. Optimized CAM-B3LYP/aug-cc-pVDZ geometries are used to generate the force field parameters with the initial solvent effects incorporated by means of the polarizable continuum model.\cite{tomasi2005} PhCN is kept frozen during the MD runs, similarly to PNA (see Ref. \citenum{ambrosetti2021quantum}). This choice is justified by their planar and rigid structure, and also avoids any potentially poor description of dihedral distributions by the classical force field.\cite{kjellgren2018importance,giovannini2018effective}
All simulations are performed using the GROMACS package.\cite{gromacs} Following a similar methodology as in Ref. \citenum{ambrosetti2021quantum}, our systems consist of a single molecule of PhCN surrounded by thousands of solvent molecules and enclosed in a simulation box of \SI{7}{\nano\meter} size. They are minimized for 500 steps, prior to a \SI{2}{\nano\second} equilibration in the isothermal-isobaric ensemble, keeping the temperature (\SI{300}{\kelvin}) and the pressure (\SI{1}{atm}) constant by means of a velocity-rescaling method,\cite{bussi2007} with a coupling constant of \SI{0.1}{\pico\second}, and the Berendsen barostat,\cite{berendsen1986practical} with a coupling constant of \SI{2.0}{\pico\second}, respectively. Values of \SI{9.7e-5}{} and \SI{9.6e-5}{\per\bar} are used for the isothermal compressibilities of THF and ACN, respectively. Afterward, an NVT production stage of \SI{10}{\nano\second} is performed in order to have a well-equilibrated system before extracting representative configurations. The LINCS algorithm\cite{hess1997lincs} is used to constrain all bonds of the solute molecule. The particle-mesh Ewald (PME) algorithm\cite{darden1993pme} is employed to handle long-range electrostatic interactions. Van der Waals and short-range electrostatic interactions are truncated with a smoothed \SI{1.4}{\nano\meter} spherical cutoff. The equations of motion are integrated with a \SI{2}{\femto\second} time step.

\begin{figure}
    \centering
    \includegraphics[width=\linewidth]{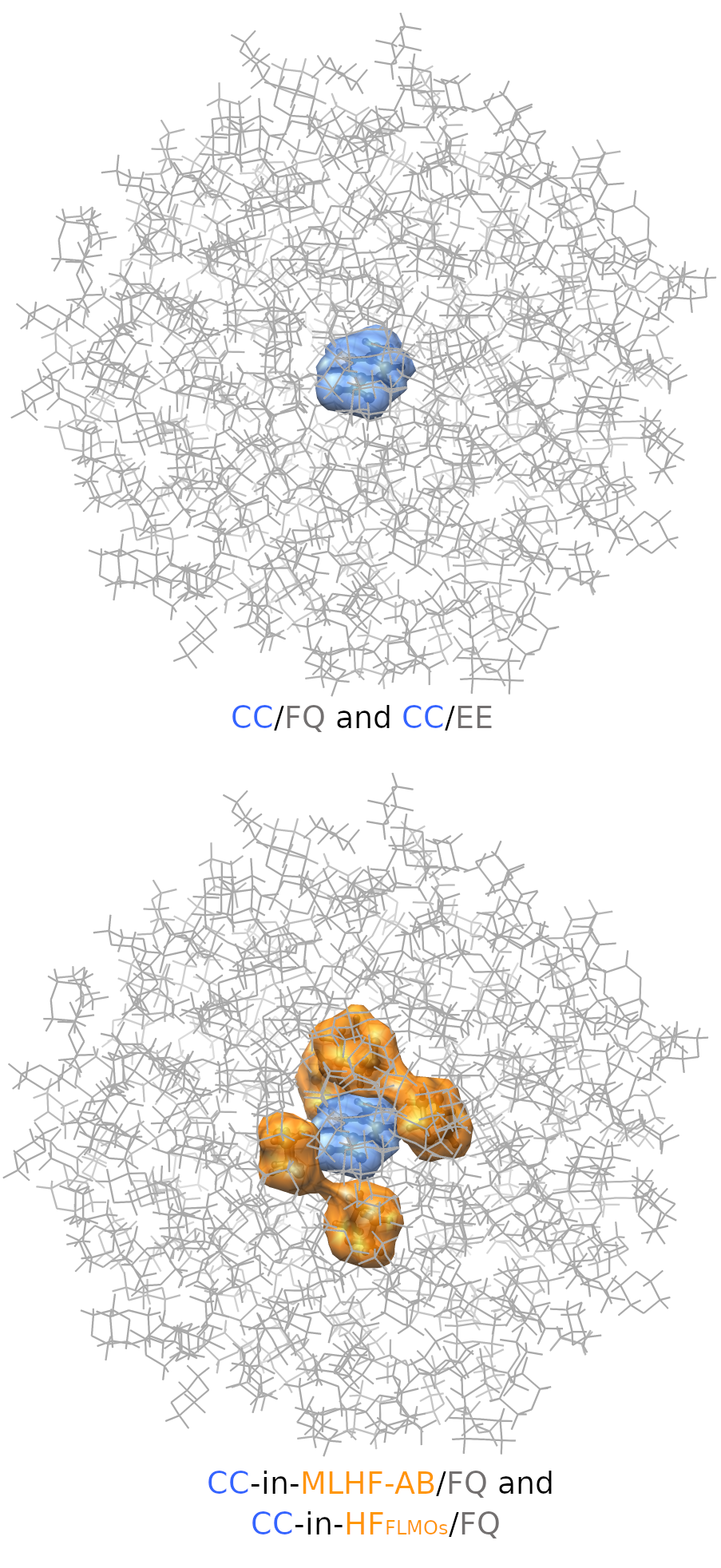}
    \caption{Graphical depiction of a snapshot of PNA-in-DIO as partitioned in the CC/FQ and CC/EE (top) and CC-in-MLHF-AB/FQ and CC-in-HF\textsubscript{FLMOs}/FQ (bottom) approaches.
    The atoms in blue are treated at the CC level, by using HF as reference wavefunction; 
    the atoms in orange are treated either at the HF level---in HF\textsubscript{FLMOs}---or as the inactive MLHF part---in MLHF-AB; the atoms in grey are treated at the MM level.}
    \label{fig:pna-in-dio_levels}
\end{figure}

A set of 20 snapshots is selected from the production stage of each MD simulation. The time separation between them (\SI{100}{\pico\second} for PNA-in-DIO and \SI{500}{\pico\second} for PhCN-in-ACN and PhCN-in-THF) is large enough to ensure that they are uncorrelated.\cite{reinholdt2018modeling, harczuk2015frequency, skoko2020simulating, puglisi2019interplay}
A droplet with a spherical shape of radius \SI{20}{\angstrom} centered on the solute is cut.
Note that the number of selected frames is enough to guarantee the convergence of the results (see Fig. S1-S3 and S14-S15 in the Electronic Supplementary Information -- ESI\dag). 
The geometries of all the frames studied in this work can be found in Ref.\citenum{geometries}.
For each extracted snapshot, the linear polarizability is then calculated by describing the whole system at different levels of theory, defined within a hierarchical ladder: (i) the solute is described at the QM level, whereas the environment is described by means of electrostatic (QM/EE) or polarizable embedding (by exploiting the FQ force field -- QM/FQ); (ii) The solute and the closest solvent molecules are included in the MLHF-AB region, while the remaining solvent molecules are described at the FQ level. The solvent molecules within a range of \SI{2.5}{\angstrom} (PNA-in-DIO), \SI{2.75}{\angstrom} (PhCN-in-THF) and \SI{3.5}{\angstrom} (PhCN-in-ACN) from any atom of the solute are included in the inactive MLHF-AB calculations, whereas the solute molecule represents the active part. Such an approach is called MLHF-AB/FQ in what follows. The same solvent molecules and the solute represent the two regions described at the HF$_{\text{FLMOs}}$ level in the HF$_{\text{FLMOs}}$/FQ approach. The partitioning of the spherical snapshots at the different levels of theory is graphically depicted in Fig. \ref{fig:pna-in-dio_levels}, by taking PNA-in-DIO as a representative example.

For MLHF-AB/FQ calculations, the protocol outlined in Ref. \citenum{goletto2021combining} is followed. A superposition of molecular densities\cite{neugebauer2005merits} is used as a starting guess. While the active MO virtual space is constructed at the beginning of the calculation by means of orthonormalized\cite{lowdin1970} PAOs,\cite{pulay1983, saebo1993} the active occupied space is firstly determined by a limited Cholesky decompositon algorithm,\cite{beebe1977simplifications, sanchez2010, aquilante2006fast, folkestad2019efficient} and then iteratively adjusted by maximizing the interaction energy $E_{AB}$ in Eq. \ref{eq:mlhf-ab_en}. The number of active MOs is selected to be equal to the correct number of occupied MOs of the active region.

In order to calculate the linear polarizability of each snapshot, the solute is described at the EOM-CC2 or EOM-CCSD level with the aug-cc-pVDZ basis set, by using the HF/EE, HF/FQ, MLHF-AB/FQ, and HF$_{\text{FLMOs}}$/FQ reference wave functions. The basis set is selected by following Refs. \citenum{egidi2014benchmark, cuesta2004polarizabilities, alparone2011theoretical, alparone2013linear}. For the MLHF-AB/FQ and HF$_{\text{FLMOs}}$/FQ reference wave functions, the solvent molecules are described with the cc-pVDZ basis set. In CC/EE, GAFF atomic charges are used for the EE region.\cite{wang2004development, wang2006automatic} For the FQ layer in CC/FQ, CC-in-MLHF-AB/FQ and CC-in-HF$_{\text{FLMOs}}$/FQ, the atomic electronegativity and chemical hardness parameters have been taken from Ref. \citenum{ambrosetti2021quantum} 

For each snapshot, we calculate the static and dynamic isotropic electronic part of the polarizabilities, which is given by

\begin{equation}
    \alpha^{\text{iso}} = \frac{1}{3} (\alpha_{xx} + \alpha_{yy} + \alpha_{zz})
\end{equation}

In the static case, a reorientation term $\alpha_\mu$ is added to the purely electronic term to yield the total static polarizability, $\alpha_{0}^{\mathrm{tot}}$:

\begin{align}
    \alpha_{0}^{\mathrm{tot}} & = \alpha^\text{iso}_0 + \alpha_\mu.\label{eq:alpha_tot_static} \\
    \alpha_{\mu} & = \frac{|\mu|^2}{3k_B T},\label{eq:alphamu}
\end{align}
where $\mu$ is the molecular dipole moment, $k_B$ is the Boltzmann constant and $T$ is the temperature.\cite{liptay1982determination, cammi2000attempt}
The final isotropic polarizability is obtained by averaging the results computed for each snapshot. It is worth noting that local field effects induced on the active part by the polarization of solvent under external radiation are not considered in this work, although they might affect computed linear response properties.\cite{list2016local,egidi2014benchmark,tomasi2002molecular} All the calculations are performed with a locally modified version of the electronic structure program $e^T$.\cite{eT_jcp} The $e^T$ default thresholds are used for the optimization of the reference and coupled cluster ground state wave functions, as well as for the dipole moments and EOM-CC polarizabilities. The threshold for the Cholesky decomposition of the two-electron repulsion integrals is set to $10^{-4}$.


\section{Numerical Results}

All the aforementioned methods are used to calculate the linear polarizabilities of PNA-in-DIO, PhCN-in-ACN and PhCN-in-THF. In this section, the computed results are analyzed in terms of the different physico-chemical solute-solvent interactions introduced by the different methods. The accuracy and robustness of the approaches are then tested by comparing the computed data with the available experimental results.\cite{wortmann1993deviations, alvarado2003solvent}

\subsection{PNA-in-DIO}

Let us discuss the case of PNA-in-DIO. PNA has been the focus of a large variety of theoretical\cite{daniel1990nonlinear, karna1991nonlinear, champagne1996vibrational,egidi2014benchmark} and experimental\cite{millefiori1977electronic, carsey1979systematics, woodford1997solvent} investigations. It is characterized by a push-pull electronic structure, presenting an electron-acceptor and an electron-donor functional groups on the opposite sides of a $\pi$-conjugated aromatic structure (see also Fig. \ref{fig:solutes-solvents}). Such a feature implies that its optical properties are strongly influenced by solvent effects,\cite{cammi2000attempt,cammi1998solvent, painelli2001linear,cammi2003multiconfigurational,giovannini2019fqfmulinear,kosenkov2010solvent} making PNA a perfect candidate for studying the performances of the different theoretical approaches.

\begin{figure}
    \centering
    \includegraphics[width=\linewidth]{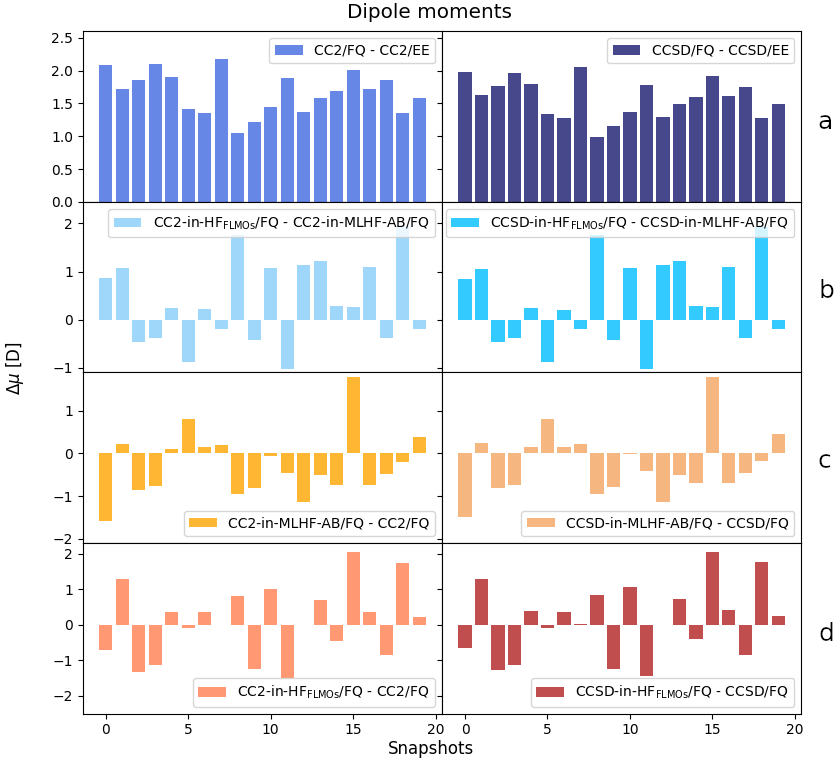}
    \caption{PNA-in-DIO snapshot-to-snapshot differences between CC-in-HF\textsubscript{FLMOs}/FQ, CC-in-MLHF-AB/FQ, CC/FQ and CC/EE ground state dipole moments results.}
    \label{fig:pna_dip}
\end{figure}

\begin{figure}
    \centering
    \includegraphics[width=\linewidth]{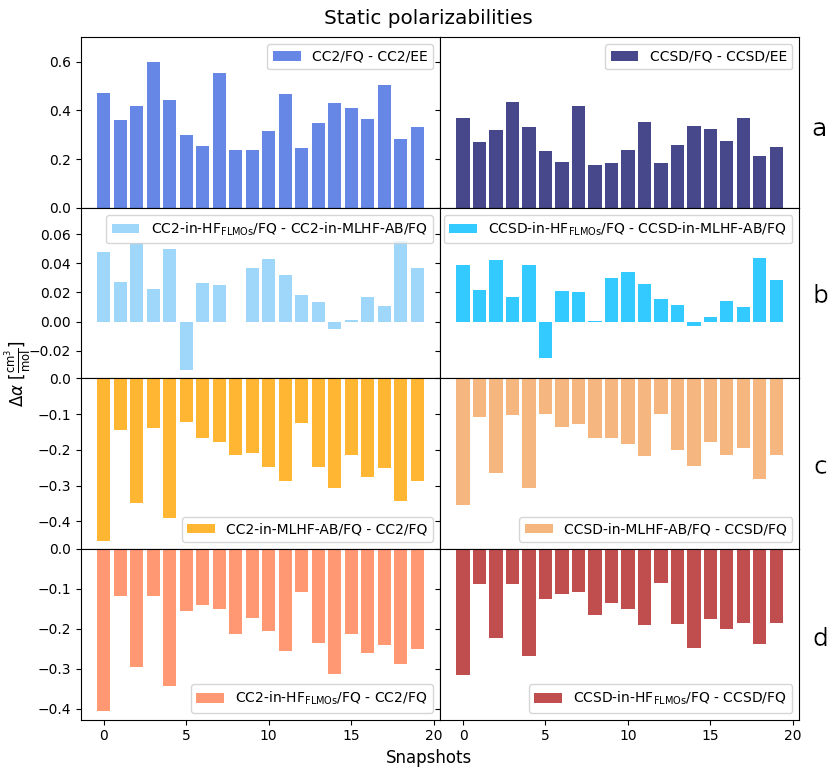}
    \caption{PNA-in-DIO snapshot-to-snapshot differences between CC-in-HF\textsubscript{FLMOs}/FQ, CC-in-MLHF-AB/FQ, CC/FQ and CC/EE results for the electronic component of the static polarizability.}
    \label{fig:pna_stat_pol}
\end{figure}

\begin{figure}
    \centering
    \includegraphics[width=\linewidth]{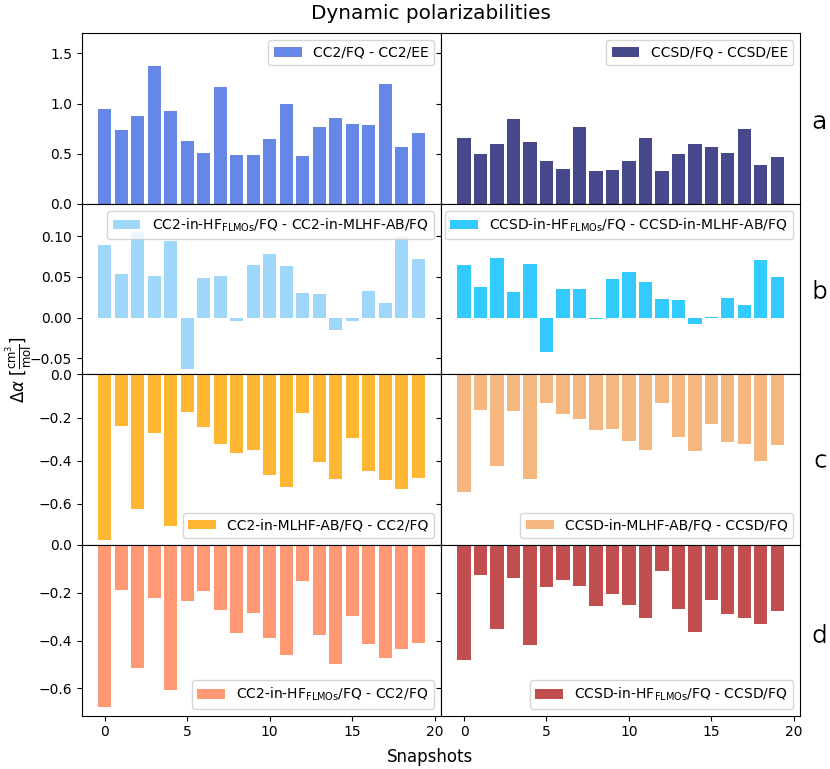}
    \caption{PNA-in-DIO snapshot-to-snapshot differences between CC-in-HF\textsubscript{FLMOs}/FQ, CC-in-MLHF-AB/FQ, CC/FQ and CC/EE electronic dynamic polarizability results (\SI{589}{nm}).}
    \label{fig:pna_dyn_pol}
\end{figure}

In order to highlight the different solute-solvent physico-chemical interactions taken into account by the different investigated approaches, we analyze the results computed for each snapshot (see Fig. \ref{fig:pna_dip}, \ref{fig:pna_stat_pol} and \ref{fig:pna_dyn_pol}). 
In Fig. \ref{fig:pna_dip}a, the differences between the dipole moments computed at the CC/FQ and CC/EE levels are graphically depicted. It can be noticed that CC/FQ dipole moments are larger than CC/EE ones, independently of the solute-solvent configurations, i.e. the snapshots. This is due to the inclusion of polarization effects, described by means of the FQ force field, which increase the magnitude of solute-solvent interactions, and consequently the computed dipole moments. Remarkably, such an increase highly varies as a function of the snapshot, ranging from 1 to about 2.2 Debye, and yields an increase of about $17$-$18\%$ on average. 

Moving to the three layers approaches, Fig. \ref{fig:pna_dip}b depicts the CC-in-HF$_{\text{FLMOs}}$/FQ and CC-in-MLHF-AB/FQ $\Delta\mu$ results as a function of the snapshot.
This case provides a different picture.
In fact, a variability both in magnitude and in sign is reported between the two approaches. However, for most snapshots, the CC-in-HF$_{\text{FLMOs}}$/FQ approach predicts larger dipole moments as compared to CC-in-MLHF-AB/FQ. Such a finding can be explained by considering that within the HF$_{\text{FLMOs}}$/FQ reference, all the solute-solvent interactions are fully accounted for at the HF level. At the MLHF-AB/FQ level, on the contrary, only part of the solute-solvent polarization effects are taken into account, and the inactive MOs---those belonging to the solvent---are not fully optimized. As a consequence, for most snapshots, the full account of polarization of CC-in-HF$_{\text{FLMOs}}$/FQ yields large dipole moments.
The negative deviations can be instead related to the optimization of the MOs of the inactive part, which enhances Pauli repulsion effects.

Finally, we compare the results obtained by exploiting CC-in-HF$_{\text{FLMOs}}$/FQ and CC-in-MLHF-AB/FQ to the CC/FQ values, as reported in Fig. \ref{fig:pna_dip}c and d, respectively. 
While CC-in-MLHF-AB/FQ generally yields a decrease of the dipole moment, a variability in intensity and sign is reported for CC-in-HF$_{\text{FLMOs}}$/FQ with respect to CC/FQ.
The numerical results can again be discussed in light of the physico-chemical interactions included in the different approaches. In the CC/FQ approach, the solute-solvent interactions are limited to electrostatics and polarization, whereas in both the three-layer methods Pauli repulsion effects are also taken into consideration. On the one hand, this explains the average decrease of the dipole moment reported for the two quantum embedding methods. On the other hand, the variability depicted in Fig. \ref{fig:pna_dip}d can suggest that for some snapshots the CC/FQ approach is not able to fully account for the solute-solvent polarization effects.

We now move to the differences in the static and dynamic polarizabilities, which are graphically depicted as a function of the snapshot in Fig. \ref{fig:pna_stat_pol} and \ref{fig:pna_dyn_pol}, respectively. 
The above discussion for the dipole moments is generally valid also for these linear-response properties, but with some noticeable exceptions. 
In fact, the differences between CC/FQ--CC/EE, and CC-in-HF$_{\text{FLMOs}}$/FQ--CC-in-MLHF-AB/FQ polarizabilities follow the same trends reported for $\mu$. 
In particular, the inclusion of polarization in CC/FQ yields an increase of both static and dynamic polarizabilities, with a larger effect on the dynamic one. 
The full account of polarization effects in HF$_{\text{FLMOs}}$ provides an overall increase of the computed properties, albeit with some negative values. 
However, the effect of using HF$_{\text{FLMOs}}$ in place of MLHF-AB is much smaller than the effect of using FQ in place of EE (on average a $0.2$-$0.4\%$ increase vs. a $4$-$7\%$ increase).
The differences between the quantum embedding models and QM/FQ (see Fig. \ref{fig:pna_stat_pol}c-d and \ref{fig:pna_dyn_pol}c-d) are negative for all the snapshots, showing that the inclusion of Pauli repulsion effects provides a general confinement of the active density. 
As a consequence, both the static and the dynamic polarizability values decrease by $2$-$3\%$, on average.

\begin{table*}
    \centering
    \begin{tabular}{l c c c c c}
    \hline
        Method & $\mu[\mathrm{D}]$ & $\alpha_{0}^{\mathrm{iso}}\big[\frac{\mathrm{cm}^3}{\mathrm{mol}}\big]$ & $\alpha^{\mu}\big[\frac{\mathrm{cm}^3}{\mathrm{mol}}\big]$  &  $\alpha_{0}^{\mathrm{tot}}\big[\frac{\mathrm{cm}^3}{\mathrm{mol}}\big]$ & $\alpha_{0}^{\mathrm{exp}}\big[\frac{\mathrm{cm}^3}{\mathrm{mol}}\big]$ \\
    \hline
        CC2 in vacuo                            & $ 6.7$ & $10.2$ & $221.7$ & $231.9$ &  \\
        CCSD in vacuo                           & $ 6.9$ & $ 9.6$ & $230.2$ & $239.9$ &\\
    \hline
        CC2/EE                                  & $ 9.1$ & $10.7$ & $403.8$ & $414.5$ & \multirow{8}{*}{$404\pm 6$}\\
        CC2/FQ                                  & $10.8$ & $11.1$ & $565.5$ & $576.6$ & \\
        CC2-in-MLHF-AB/FQ                       & $10.5$ & $10.8$ & $536.4$ & $547.2$ & \\
        CC2-in-HF\textsubscript{FLMOs}/FQ       & $10.8$ & $10.8$ & $573.9$ & $584.8$ & \\
        CCSD/EE                                 & $ 9.1$ & $10.0$ & $406.1$ & $416.1$ &\\
        CCSD/FQ                                 & $10.7$ & $10.3$ & $558.6$ & $568.9$ &\\
        CCSD-in-MLHF-AB/FQ                      & $10.4$ & $10.1$ & $532.4$ & $542.5$ &\\
        CCSD-in-HF\textsubscript{FLMOs}/FQ      & $10.8$ & $10.1$ & $569.4$ & $579.5$ & \\
    \hline
    \end{tabular}
    \caption{Calculated PNA-in-DIO dipole moments ($\mu$), isotropic electronic static polarizabilities ($\alpha_{0}^{\text{iso}}$),  reorientation terms ($\alpha^\mu$), and total static polarizabilities ($\alpha_{0}^\text{tot}$). The experimental reference\cite{wortmann1993deviations} is also provided, along with CC2 and CCSD \textit{in vacuo} values.}
    \label{tab:pna_static}
\end{table*}


\begin{table}
    \centering
    \begin{tabular}{l c c}
    \hline
        Method & $\alpha_{\omega}^{\mathrm{iso}}\big[\frac{\mathrm{cm}^3}{\mathrm{mol}}\big]$ & $\alpha_{\omega}^{\mathrm{exp}}\big[\frac{\mathrm{cm}^3}{\mathrm{mol}}\big]$ \\
    \hline
        CC2 in vacuo                       & $11.0$ & \\
        CCSD in vacuo                      & $10.3$ & \\
    \hline
        CC2/EE                             & $11.9 $ & \multirow{8}{*}{$14.1 \pm 0.4$
        } \\
        CC2/FQ                             & $12.6 $ & \\
        CC2-in-MLHF-AB/FQ                  & $12.2 $ & \\
        CC2-in-HF\textsubscript{FLMOs}/FQ  & $12.3 $ &\\
        CCSD/EE                            & $10.9 $ & \\
        CCSD/FQ                            & $11.5 $ & \\
        CCSD-in-MLHF-AB/FQ                 & $11.2 $ & \\
        CCSD-in-HF\textsubscript{FLMOs}/FQ & $11.2 $ &\\
    \hline
    \end{tabular}
    \caption{Calculated PNA-in-DIO dynamic polarizabilities (\SI{589}{nm}). The experimental reference\cite{wortmann1993deviations} is also provided, along with CC2 and CCSD \textit{in vacuo} values.}
    \label{tab:pna_dyn}
\end{table}

Finally, let us move to comment on the averaged results, which can be compared to the available experimental data. In Table \ref{tab:pna_static} the averaged isotropic values of both the electronic static polarizability ($\alpha_0^{\text{iso}}$) and the dipole moment obtained with the different theoretical methods are reported, together with the experimental value from Ref. \citenum{wortmann1993deviations}. To better quantify the solvent effects on the computed properties, Table \ref{tab:pna_static} also lists the \textit{vacuo} values calculated at the CC2 and CCSD levels. The total static polarizability, $\alpha_{0}^{\text{tot}}$, is computed using Eq.~\ref{eq:alpha_tot_static}, so it consists of an orientationally averaged polarizability ($\alpha_\mu$) and of the purely electronic contribution $\alpha^{\text{iso}}_0$. 

The CC2 and CCSD results mainly differ in two respects: while CC2 $\alpha^{\text{iso}}_{0}$ is larger than the CCSD counterpart, the opposite holds for the dipole moments in the gas phase. Solvent effects can be appreciated by comparing the gas-phase results with those obtained by including a description of the embedding. In particular, an overall increase of both the static polarizability ($4$-$8\%$) and the dipole moments ($33$-$61\%$) is highlighted. Such a trend is compatible with what has been reported in the literature for similar systems.\cite{mikkelsen1994solvent, cammi2003multiconfigurational, targema2013molecular} While the CC/EE approaches provide the smallest values of $\mu$ ($\sim$ \SI{1.5}{D} away from all other methods) and $\alpha^{\text{iso}}_0$, QM/FQ reports the largest $\alpha^{\text{iso}}_0$, and CC-in-HF$_{\text{FLMOs}}$/FQ the largest $\mu$. All the trends between the values computed with different solvation approaches follow what has already been pointed out for Fig. \ref{fig:pna_dip}--\ref{fig:pna_dyn_pol}. Indeed, CC/FQ $\mu$ results are larger than CC/EE $\mu$, and a similar trend is reported for CC-in-HF\textsubscript{FLMOs}/FQ as compared to CC-in-MLHF-AB/FQ. Such a result is primarily due to the inclusion (in CC/FQ) and full accounting (in CC-in-HF\textsubscript{FLMOs}/FQ) of polarization. 
When considering $\alpha^{\text{iso}}_{0}$, the aforementioned trends remain valid.
In addition, we note that the use of the multilevel wave function as a reference decreases $\alpha^{\text{iso}}_{0}$ with respect to the CC/MM data.
This is in agreement with Fig. \ref{fig:pna_stat_pol} and \ref{fig:pna_dyn_pol}, and is due to the quantum confinement of the solute density as a result of the solute-solvent Pauli repulsion contributions introduced by the multilevel modeling. The differences between CC2 and CCSD dipole moment results are small, regardless of the solvation model employed.

When comparing to the experimental reference in Ref.~\citenum{wortmann1993deviations} (see Table \ref{tab:pna_static}), it is worth remarking that the largest contribution to $\alpha^{tot}_0$ is given by $\alpha^{\mu} \propto \mu^2$. 
Hence, the computed values of the total static polarizability strongly depend on the numerical values of $\mu$. Indeed, CC/EE reports the smallest $\alpha^{tot}_0$, whereas the largest $\alpha^{tot}_0$ values are given by CC-in-HF\textsubscript{FLMOs}/FQ, which has a similar performance as CC-in-MLHB-AB/FQ, and CC/FQ. The differences between CC/EE and the other solvation methods range between $34$ and $45\%$. This is primarily due to the differences in the computed dipole moments, as the $\alpha^{\text{iso}}_0$ only slightly affects the final computed property. 
Since CC/EE $\alpha^{tot}_0$ is closest to the experimental value, the most important error source for the other methods lies in the overestimation of $\mu$, which is enhanced when the polarization is included in the modeling, and lowered by the Pauli repulsion effects introduced in multilevel methods. 
A small difference in $\mu$ is reflected in large differences in $\alpha^\mu$, and consequently in $\alpha_{0}^\text{tot}$.  In fact, as commented above, $\alpha_{0}^\text{tot}$ is calculated as the sum of $\alpha_{0}^\text{iso}$ and $\alpha^\mu$, and our results show that $\alpha^\mu$ is approximately 50 times larger than $\alpha_0^\text{iso}$.

To remove the dependency of the results on the computed dipole moments, we move to dynamic polarizabilities, for which $\alpha_\mu = 0$. Therefore, the computed $\alpha^{\text{iso}}_\omega$ can be directly compared to the experimental values, which are given in Table \ref{tab:pna_dyn}. Similar trends as reported for the static electronic polarizability (see Table \ref{tab:pna_static}) can be observed also for the frequency-dependent case. The inclusion of solvent effects in the modeling increases the computed values independently of the exploited method, ranging between a $6\%$ (CCSD/EE) and a $15\%$ (CC2/FQ) shift. The gap between CCSD and CC2 results is larger as compared to the static polarizability, with the CCSD results being lower than CC2 by up to $9\%$ (CC/FQ). The trends between the different approaches directly follow those discussed in the static case: CC/FQ $\alpha^{\text{iso}}_{\omega}$ are larger than the corresponding CC/EE values, and the same is generally valid between CC-in-HF$_{\text{FLMOs}}$/FQ and CC-in-MLHF-AB/FQ, due to the inclusion and the full accounting of polarization effects, respectively. Similarly to the static case, CC-in-HF$_{\text{FLMOs}}$/FQ and CC-in-MLHF-AB provide very similar results, indicating that the MLHF-AB/FQ method is able to account for most of the polarization effects.  On the contrary, by moving from CC/FQ to the multilevel methods the value of the computed property decreases ($\sim 2$-$3\%$), because the Pauli repulsion yields a confinement effect in the reference wave function.

Comparing the results in Table \ref{tab:pna_dyn} with the experimental counterpart, we first note that, differently from the static case, all the computed values are smaller. The CC/EE results present the largest deviations, while the CC/FQ ones are closest to experiment. As commented above, CC2 polarizabilities are generally larger than CCSD ones, thus resulting in a better agreement with the experimental reference. However, it is worth noting that CC2 might reportedly overestimate linear polarizabilities,\cite{christiansen1999frequency, salek2005comparison} probably due to an overestimation of the dispersion interaction.\cite{rocha2009linear, schmies2011structures, kolaski2013aromatic} Considering the high level of theory employed in this work, the systematic underestimation of all methods could be explained by the fact that our model discards the zero-point correction and the pure vibrational contribution to the linear response properties. While the latter plays a negligible role in determining dynamic polarizabilities (due to the unfavourable dependence on the external frequency), studies at the DFT level have shown that the former can increase the purely electronic contribution by up to $6\%$.\cite{egidi2014stereoelectronic} Considering that the discrepancies with the experimental counterpart range between $1.5$ ($\sim 10\%$, with CC2/FQ) and $3.2$ ($\sim 22\%$, with CCSD/EE) cm$^3$/mol, an overall agreement with the experiment can be reported for almost all methods. Finally, it is worth noting that our modeling neglects explicit terms arising from polarization contributions in response equations, which may enhance the computed linear polarizabilities, as reported in similar contexts.\cite{cammi2012coupled, caricato2018coupled, caricato2019coupled, caricato2020coupled}


\subsection{PhCN-in-ACN and PhCN-in-THF}

We now move to the case of PhCN-in-ACN and PhCN-in-THF polarizabilities, which have been studied from the experimental point of view in Ref. \citenum{alvarado2003solvent} as a function of the external frequency (ranging from \SI{1.553}{} to \SI{2.295}{\per\micro\meter}). We again model solvent effects by means of CC/EE, CC/FQ and CC-in-MLHF-AB/FQ; the CC-in-HF\textsubscript{FLMOs}/FQ methods have not been included in the comparison, as in the previous section the differences with CC-in-MLHF-AB/FQ in the polarizabilities have been found to be negligible, but the method is associated with a higher computational cost. 

\begin{figure}
    \centering
    \includegraphics[width=\linewidth]{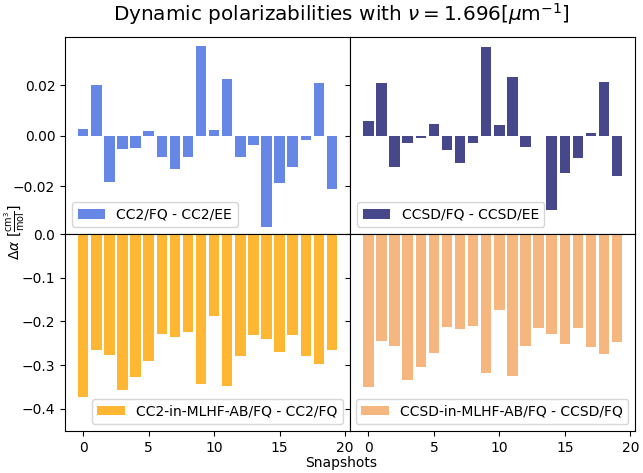}
    \caption{PhCN-in-ACN snapshot-to-snapshot differences between CC-in-MLHF-AB/FQ, CC/FQ and CC/EE electronic dynamic polarizability results (\SI{1.696}{\per\micro\meter}).}
    \label{fig:bnz-in-acn_bars}
\end{figure}

\begin{figure}
    \centering
    \includegraphics[width=\linewidth]{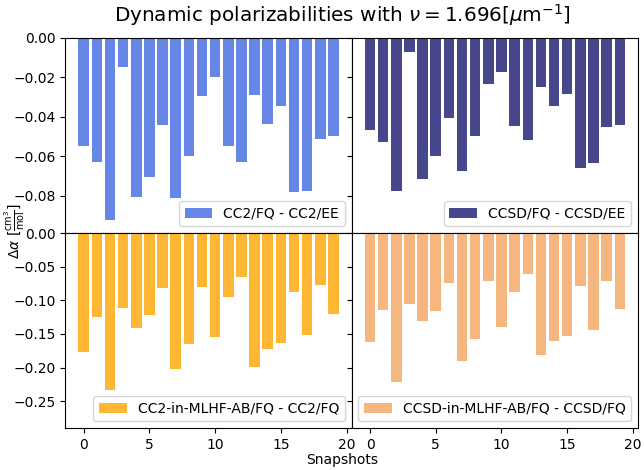}
    \caption{PhCN-in-THF snapshot-to-snapshot differences between CC-in-MLHF-AB/FQ, CC/FQ and CC/EE electronic dynamic polarizability results (\SI{1.696}{\per\micro\meter}).}
    \label{fig:bnz-in-thf_bars}
\end{figure}

The snapshot-to-snapshot differences CC/FQ--CC/EE and CC-in-MLHF-AB/FQ--CC/FQ are depicted in Fig. \ref{fig:bnz-in-acn_bars} and \ref{fig:bnz-in-thf_bars} for the PhCN-in-ACN and PhCN-in-THF systems, respectively. The differences are presented for a specific frequency equal to \SI{1.696}{\per\micro\m}, which corresponds to the experimental sodium D line. The plots for all the other frequencies considered in this work are reported in Fig. S4-S13 of the ESI\dag, and show similar trends. 

For PhCN-in-ACN (see Fig. \ref{fig:bnz-in-acn_bars}), the differences between CC/FQ and CC/EE are negligible, ranging between $\pm$ 0.02 cm$^3$/mol. Additionally, the differences strongly depend on the specific solute-solvent configurations and display a sign alternation, thus averaging out in the final property. Note that, in contrast to what has been discussed in the previous section (see Fig. \ref{fig:pna_dyn_pol}), the sign alternation indicates that the parametrization of the EE modeling overestimates electrostatics, similarly to other non-polarizable force fields.\cite{mobley2007comparison} The mutual polarization between PhCN and ACN appears to play a minor role in the solute-solvent interaction, as compared to PNA-in-DIO system. On the other hand, the introduction of the intermediate MLHF-AB layer between the coupled cluster and FQ regions lowers the polarizabilities results in all the snapshots, with differences ranging from 0.2 to 0.4 cm$^3$/mol. This indicates once again the confinement effects provided by the accounting for solute-solvent Pauli repulsion.

For PhCN-in-THF, the CC/FQ values are smaller than the CC/EE values in all snapshots. Such a finding is opposite to the PNA-in-DIO case, and shows that the parametrization exploited in the EE force field is including (overestimating) electrostatic effects. However, the CC/FQ--CC/EE differences only reach the $1\%$ of the total value of the polarizability. Therefore the THF polarization, albeit numerically more significant than that of ACN (see Fig. \ref{fig:bnz-in-acn_bars}), does not play a significant role for such systems. On the contrary, the Pauli repulsion interactions introduced by the MLHF-AB layer have again a larger influence on the polarizabilities, lowering the results by approximately $3\%$ (and numerically by 0.1--0.25 cm$^3$/mol).

\begin{figure}
    \centering
    \includegraphics[width=\linewidth]{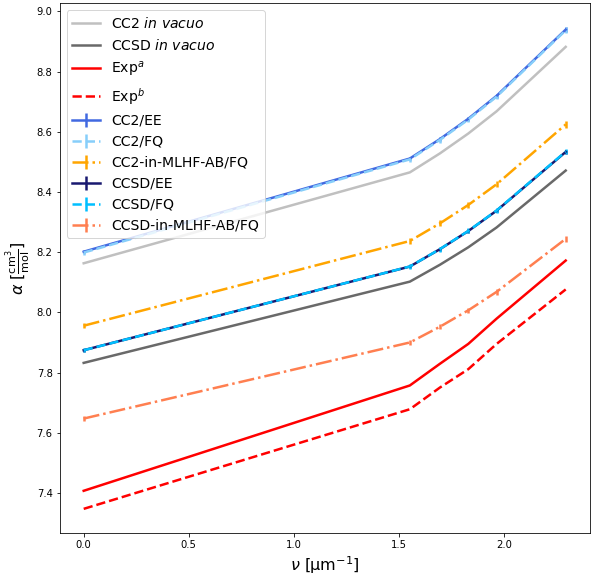}
    \caption{Calculated PhCN-in-ACN polarizabilities as a function of the external frequency. Standard error bars at the 68\% confidence interval are plotted. PhCN \textit{in vacuo} polarizabilities are also given, along with the experimental benchmark from Ref. \citenum{alvarado2003solvent}. $\text{Exp}^a$ employs the Lorentz local field correction, while $\text{Exp}^b$ employs the Onsager local field correction.}
    \label{fig:bnz-in-acn_plots}
\end{figure}

\begin{figure}
    \centering
    \includegraphics[width=\linewidth]{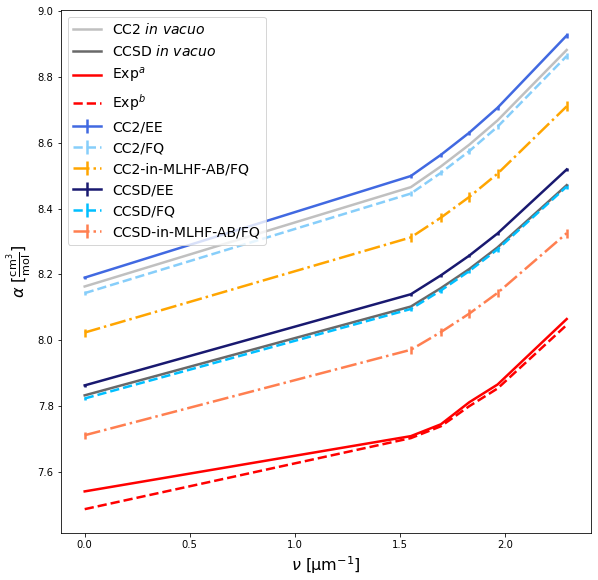}
    \caption{Calculated PhCN-in-THF polarizabilities as a function of the external frequency. Standard error bars at the 68\% confidence interval are plotted. PhCN \textit{in vacuo} polarizabilities are also given, along with the experimental benchmark from Ref. \citenum{alvarado2003solvent}. $\text{Exp}^a$ employs the Lorentz local field correction, while $\text{Exp}^b$ employs the Onsager local field correction.}
    \label{fig:bnz-in-thf_plots}
\end{figure}

The computed averaged values of the PhCN-in-ACN static/dynamic polarizabilities are plotted as a function of the external frequency in Fig. \ref{fig:bnz-in-acn_plots} (see Table S1 in the ESI$\dag$ for the numerical data), together with the \textit{in vacuo} data and the experimental results reproduced from Ref. \citenum{alvarado2003solvent}. In particular, we report two different experimental references, which are obtained by applying the Lorentz ($\text{Exp}^a$) and Onsager ($\text{Exp}^b$) local field corrections to the measured refractive indexes. Indeed, it is worth remarking that the reported data are not the measured quantities---that is, the refractive indexes---, but the electronic part of the polarizability extracted from them. Therefore, the reference data are associated with an intrinsic systematic error related to the approach exploited to extrapolate a microscopic quantity (the polarizability) from a macroscopic one (the refractive index). 

In Fig. \ref{fig:bnz-in-acn_plots}, the averaged polarizabilities computed by all the different methods are shown to follow the same trend with respect to the external frequency. In particular, CC/EE and CC/FQ results are almost identical, whereas the inclusion of the MLHF-AB layer lowers the polarizabilities by approximately $3\%$. Indeed, with respect to the \textit{in vacuo} results, CC/EE and CC/FQ yield an increase in the polarizabilities, albeit with a negligible deviation ($< 1\%$). On the other hand, the values obtained at the CC-in-MLHF-AB/FQ level are $2$ to $3\%$ lower than the corresponding CC results \textit{in vacuo}. Thus, depending on how the environment is treated, solvent effects shift the polarizabilities to opposite directions. As the main difference between the approaches lies in the Pauli repulsion between PhCN and ACN being mostly taken into account in CC-in-MLHF-AB/FQ, this contribution appears to have a significant influence on the results. CCSD polarizabilities are approximately $4\%$ smaller than the corresponding CC2 values with all the multiscale methods, following the same trend observed \textit{in vacuo}.

We now move to the comparison with the experimental reference.\cite{alvarado2003solvent} Regardless of the choice of local field correction, being it Lorentz or Onsager, the experimental results are lower than the computed values. In particular, $\text{Exp}^b$ and  $\text{Exp}^a$ polarizabilities differ of about $1\%$, the former being the lowest. The best agreement with the experiment is obtained by using CCSD-in-MLHF-AB/FQ, which presents $1$-$3\%$ and $2$-$4\%$ deviations from $\text{Exp}^a$ and $\text{Exp}^b$, respectively. In particular, CCSD-in-MLHF-AB/FQ reports its largest discrepancy with the experiment for the static polarizability. Here, the values in Ref. \citenum{alvarado2003solvent} are not recovered from the experimental permittivity, but obtained by extrapolation with a Cauchy-type dispersion curve fit, which might introduce further inaccuracy. The worst agreement with the experiment is given by CC2/EE and CC2/FQ, which deviate from the experiment by $9\%$ and $12\%$, respectively. Remarkably, among the considered methods, only the three-layer CC-in-MLHF-AB/FQ approaches shift the computed polarizabilities from the corresponding \textit{in vacuo} results towards the experiment.

Fig. \ref{fig:bnz-in-thf_plots} depicts the averaged polarizability values (see Table S2 in the ESI$\dag$ for the numerical data), the \textit{in vacuo} reference, and experimental benchmark\cite{alvarado2003solvent} for the PhCN-in-THF system, plotted again as a function of the external frequencies. Similarly to the previous case, CC/EE and CC/FQ results are very similar, with CC/FQ providing computed polarizabilities less than $1\%$ smaller than CC/EE. This is in agreement with the results reported in Fig. \ref{fig:bnz-in-thf_bars}. An additional decrease is given by CC-in-MLHF-AB/FQ, with values $1$-$2\%$ smaller than CC/FQ. The CC2 and CCSD values show the same trend with respect to a different treatment of the environment, with CCSD providing polarizabilities $4$-$5\%$ smaller than CC2. The same behaviour is observed for the \textit{in vacuo} results. When comparing to the \textit{in vacuo} reference, the CC/FQ values are almost identical, with deviations $<0.3\%$. While being very similar, the CC/EE polarizabilities are slightly larger, even if the differences fall below $1\%$. On the other hand, the CC-in-MLHF-AB/FQ methods lower the polarizabilities by $2\%$ with respect to the \textit{in vacuo} results.

Additionally, Fig. \ref{fig:bnz-in-thf_plots} shows that the experimental references are once again smaller than the computed values. The comparison between the methods follows the same trends as in PhCN-in-ACN, with CCSD-in-MLHF-AB/FQ being the closest (with a $2$-$4\%$ deviation) and CC2/EE being the furthest (with a $9$-$11\%$ deviation) from the experiment. While starting from the \textit{in vacuo} reference CC/EE goes in the wrong direction with respect to the experiment, both CC/FQ and CC-in-MLHF-AB/FQ lower the polarizabilities towards the experiment.


\section{Summary and conclusions}

We have presented a computational investigation of linear polarizabilities of organic moieties embedded in non-aqueous solvents, employing different strategies to model solvent effects. We have considered a hierarchy of solvation approaches, ranging from common QM/MM methods to three-layer approaches based on a multilevel partitioning of the reference wave function. In particular, we have considered both non-polarizable and polarizable QM/MM approaches, the latter based on the FQ force field, suitably parameterized for the selected solvents. The three-layer approaches are instead based on a partitioning of the system into three portions: an active region (the solute), an inactive region (the solvent molecules closest to the solute), and an MM region (the rest of the solvent), treated by means of the FQ force field. In this way, the external MM layer accounts for long range interactions with an electrostatic and polarization nature. The MLHF-AB reference wave function introduces the electron repulsion effects between solute and solvent, as well as an approximated HF treatment of the polarization. 

The approaches have been applied to the calculation of static and dynamic linear polarizabilities (at the CC2 and CCSD levels) of the PNA-in-DIO, PhCN-in-ACN, and PhCN-in-THF systems. To sample the solute-solvent phase-space, the calculations have been run on various snapshots extracted from classical MD simulations. Overall, this protocol gives a consistent physical description of the properties and interactions at work in a complex environment. The computed results have been rationalized in terms of the different solute-solvent physico-chemical interactions modeled by each solvation approach and compared with the available experimental data. In all cases, we have obtained an overall good agreement with the reference measurements, in particular when Pauli repulsion effects, which are introduced in the three-layer approaches, are taken into account. 

To further increase the agreement with experimental results, the three-layer model could be further improved to include dispersion effects between the solute and the solvent. Also, the quality of long-range electrostatics and polarization effects can be increased by including an additional source of polarization in the FQ force field in terms of fluctuating dipoles (FQF$\mu$) to account for anisotropic interactions.\cite{giovannini2019fqfmu,giovannini2019fqfmuir,marrazzini2020calculation} Finally, the protocol can be extended to the treatment of higher-order properties, such as first-hyperpolarizabilities ($\beta$).


\section*{Conflicts of interest}
There are no conflicts to declare.

\section*{Acknowledgements}
We thank Eirik F. Kjønstad, Sarai D. Folkestad and Alexander C. Paul for their contributions to the $e^T$ code. J.H.A. acknowledges Sonia Coriani (DTU) for discussions. We acknowledge funding from the Marie Sklodowska-Curie Interational Training Network “COSINE - COmputational Spectroscopy In Natural sciences and Engineering”, Grant Agreement No. 765739, and from the Research
Council of Norway through the FRINATEK project 275506, TheoLight. We acknowledge computing resources through UNINETT Sigma2---the National Infrastructure for High Performance Computing and Data Storage in Norway, through project number NN2962k.
We also acknowledge Chiara Cappelli (SNS) for computing resources, and the Center for High Performance Computing (CHPC) at SNS for providing the computational infrastructure.


\balance

\bibliography{polar.bib,biblio.bib} 
\bibliographystyle{rsc}

\end{document}